\documentclass[twocolumn,trackchanges,times]{aastex631}

\newcommand*{\teff}{$T_{\rm eff}$}
\newcommand*{\logg}{$\log~g$}

\newcommand*{\afe}{[$\alpha$/Fe]}

\newcommand*{\kms}{km s$^{-1}$}

\newcommand*{\rmax}{$r_{\rm max}$}
\newcommand*{\rmin}{$r_{\rm min}$}

\newcommand*{\ly}{$L_{\rm Y}$}
\newcommand*{\lz}{$L_{\rm Z}$}
\newcommand*{\vgsr}{$V_{\rm GSR}$}
\newcommand{\RomanNumeralCaps}[1] 
    {\MakeUppercase{\romannumeral #1}} 

\usepackage{amsmath}
\usepackage{url}
\usepackage{natbib}
\usepackage{tabularx}
\usepackage{color}
\usepackage{graphicx}
\usepackage{epstopdf}
\usepackage{gensymb}
\usepackage{hyperref}
\usepackage{longtable}
\usepackage{subfigure}
\usepackage[T1]{fontenc}
\DeclareUnicodeCharacter{2212}{-}

\begin{document}

\shorttitle{A Dynamically Distinct Stellar Population in the Sgr Stream}
\shortauthors{Kang et al.}

\title{A Dynamically Distinct Stellar Population in the Leading Arm of the Sagittarius Stream}
\author{Gwibong Kang}
\affiliation{Department of Astronomy, Space Science, and Geology, Chungnam National University, Daejeon 34134, Republic of Korea}
\author[0000-0001-5297-4518]{Young Sun Lee}
\affiliation{Department of Astronomy and Space Science, Chungnam National University, Daejeon 34134, Republic of Korea; youngsun@cnu.ac.kr}
\affiliation{Department of Physics and Astronomy and JINA Center for the Evolution of the Elements, University of Notre Dame, IN 46556, USA}
\author[0000-0002-6411-5857]{Young Kwang Kim}
\affiliation{Department of Astronomy and Space Science, Chungnam National University, Daejeon 34134, Republic of Korea}
\author[0000-0003-4573-6233]{Timothy C. Beers}
\affiliation{Department of Physics and Astronomy and JINA Center for the Evolution of the Elements, University of Notre Dame, IN 46556, USA}

\begin{abstract}

We present a chemical and dynamical analysis of the leading arm (LA) and
trailing arm (TA) of the Sagittarius (Sgr) stream, as well as for
the Sgr dwarf galaxy core (SC), using red giant branch, main sequence,
and RR Lyrae stars from large spectroscopic survey data. The different
chemical properties among the LA, TA, and SC generally
agree with recent studies, and can be understood by radial metallicity
gradient established in the progenitor of the Sgr dwarf, followed by
preferential stellar stripping from the outer part of the Sgr progenitor.
{One striking finding is a relatively larger fraction
of low-eccentricity stars ($e < 0.4$) in the LA than in the TA and SC.
The TA and SC exhibit very similar distributions.}
Considering that a tidal tail stripped off from
a dwarf galaxy maintains the orbital properties of its progenitor,
we expect that the $e$-distribution of the LA should be similar to that of the
TA and SC. Thus, the disparate behavior of the $e$-distribution of the LA
is of particular interest. {Following the analysis of Vasiliev et al.,
we attempt to explain the different $e$-distribution by introducing a time-dependent
perturbation of the Milky Way by the Large Magellanic Cloud (LMC)'s gravitational
pull, resulting in substantial evolution of the angular momentum of the LA stars
to produce the low-$e$ stars. In addition, we confirm from RR Lyrae stars
with high eccentricity ($e >$ 0.6) that the TA stars farther away
from the SC are also affected by disturbances from the LMC.}
\end{abstract}


\section{Introduction}

Numerous studies have argued that some or all of the halo system of the Milky Way (MW)
has formed through multiple mergers with dwarf galaxies of various masses \citep[e.g.,][]{bullock2005,cooper2010},
likely over an extended period of time.
An excellent example of one such merging episode is the Sagittarius (Sgr)
dwarf spheroidal galaxy, which is in the process of being
tidally disrupted \citep{ibata1994}. The Sgr dwarf was discovered as an
over-density of stars with distinctive
velocities and positions in the sky \citep{ibata1994,majewski2003,belokurov2006}.
Subsequently, the Sgr tidal stream was discovered \citep{yanny2000,ibata2001}, and its detailed
shape was later confirmed \citep{newberg2002,majewski2003}.
The Sgr stellar stream appears to wrap around the MW two
or three times \citep[e.g.,][]{majewski2003,law2010,belokurov2014},
and consists of two arms, the leading and trailing arms.

Recent studies have identified a significant metallicity discrepancy
between the leading and trailing arms, with the former being more metal-poor
by about 0.3 dex \citep{carlin2018,li2019,hayes2020,ramos2022}.
In addition, a metallicity gradient has been detected
not only in the Sgr core, but also in the Sgr
stream \citep{carlin2012,gibbons2017,hayes2020,ramos2022}.
These chemical patterns imply the existence of a significant metallicity gradient
in the Sgr progenitor galaxy. The observed metallicity difference between
the two arms and the core is thought to be the result of the selective stripping
of older, more metal-poor stars located in the outermost regions of the Sgr progenitor.
It has also been reported that, within each arm, the metal-poor stars exhibit
larger velocity dispersions than the relatively more metal-rich
stars \citep{gibbons2017,johnson2020,limberg2023}.

This overall picture has become even more complicated, with the discovery of
bifurcation in each arm \citep{belokurov2006,koposov2012},
requiring a more detailed analysis of each arm. In addition, \citet{vasiliev2021} (hereafter; V21)
reported a misalignment between the Sgr stream track and
the motion of the leading-arm stars {(corrected
for the Solar reflex motion in their observed proper motions). They explained such
a misalignment by introducing the time-dependent perturbation of the MW caused
by the Large Magellanic Cloud (LMC).}
Their finding demonstrated that the role of the LMC should be taken into account
for a better understanding of the dynamical properties and evolution of the Sgr stream.

In this letter, we employ giant branch (RGB), main
sequence (MS), and RR Lyrae stars, to identify the presence of a large fraction
of low-eccentricity ($e <$ 0.4) stars in the leading arm, as compared to both
the trailing arm and the Sgr core, which we take as a tell-tale sign for the dynamical evolution of the
leading arm due to time-dependent perturbations.

Section \ref{sec2} describes the spectroscopic survey data
from which members of the Sgr stream and core are obtained, along with
calculations of their orbital parameters. Section \ref{sec3} outlines
how we select the genuine Sgr member stars. In Section \ref{sec4}, we present
abundance and dynamical characteristics of the Sgr members, and
discuss the possible origins of low-eccentricity stars in the leading arm.
Our conclusions are presented in Section \ref{sec5}.

\section{Initial Samples} \label{sec2}
\subsection{Large Spectroscopic Survey Data} \label{sec2.1}

To gather large numbers of stars useful for the selection of members in the Sgr core
and stream, we combined various large spectroscopic survey data from the Sloan Digital
Sky Survey (SDSS; \citealt{york2000}) and its various sub-surveys, in particular from
the Sloan Extension for Galactic Understanding and Exploration \citep[SEGUE;][]{yanny2009,rockosi2022}
and the Apache Point Observatory Galactic Evolution Experiment Data Release 17 \citep[APOGEE DR17;][]{abdurro'uf2022},
as well as the Large sky Area Multi-Object Fiber Spectroscopic Telescope Data Release 5 \citep[LAMOST DR5;][]{luo2019}.
We obtained stellar parameters such as effective
temperature (\teff), surface gravity (\logg), and metallicity ([Fe/H]) by processing
the SDSS and LAMOST spectra with signal-to-noise ratios (S/Ns) larger than 10
through the SEGUE Stellar Parameter Pipeline \citep[SSPP;][]{lee2008a,lee2008b,lee2011a}.
The interested reader is referred to \citet{lee2015} for the application of the SSPP on the LAMOST spectra.
Note that the SSPP also produces estimates of [Mg/Fe] for both of the
lower-resolution spectroscopic surveys.
In order to ensure reliable chemical-abundance ratios of APOGEE stars, which
were derived from the APOGEE Stellar Parameters and Chemical Abundances
Pipeline \citep[ASPCAP;][]{garcia2016}, we only selected stars
with S/N $>$ 50 and \logg\ $<$ 4.3 from APOGEE DR17.

As reported by \citet{lee2023}, the stellar parameters and abundances for the SDSS
and LAMOST stars agree very well; {the mean
difference is about 0.1 dex for [Fe/H] and 0.01 dex for [Mg/Fe].
Similarly, we found a mean offset of --0.04 dex for [Fe/H]
and --0.02 dex for [Mg/Fe], using stars in common between SDSS/LAMOST and APOGEE.
Because the systematic offsets between surveys
are less than the typical uncertainties ($\sim$ 0.2 dex for [Fe/H] and
$\sim$ 0.1 dex for [Mg/Fe]) we did not adjust for the small mean offsets.}
We placed the radial velocity (RV) of SDSS/LAMOST stars to
the scale of the RVs reported in Gaia Early Data Release \citep[EDR3;][]{gaia2021},
using the stars in common, as described in \citet{lee2023}.

{The distance estimates for the SDSS stars was obtained by employing the
method of \citet{beers2000,beers2012}; its reported uncertainty is on the
order of 15 -- 20\%, while that of the LAMOST stars was taken from the
value-added catalog of LAMOST DR5 \citep{xiang2019}. The distances in this catalog were
estimated based on a Bayesian approach \citep{wang2016}; its uncertainty is about 20\%.
We employed the AstroNN spectro-photometric
distance obtained from a deep neural network approach for the APOGEE
stars \citep{leung2019}. The quoted uncertainty is
less than 10\%.}

In this study, we selected bright RGB and MS stars, which
can probe up to the distance of the Sgr core and stream by
imposing the following cuts: 0.3 $\leq$ $(g-r)$ $\leq$ 1.2,
\logg\ $<$ 3.5, and 4400\,K $\leq$ \teff\ $\leq$ 5600\,K for the RGB stars.
For the MS stars, cuts of 0.0 $\leq$ $(g-r)$ $\leq$ 1.2,
\logg\ $\geq$ 3.5, and 4400\,K $\leq$ \teff\ $\leq$ 7000\,K were adopted.

Note that we have removed stars located in or near known globular clusters and
open clusters. For multiply observed stars, we chose to
include the star with highest S/N, or the one observed in APOGEE if a star was observed
in both APOGEE and SDSS/LAMOST. Upon applying these various cuts, we obtained a total
of approximately 2.9 million stars from SDSS/LAMOST and about 630,000 stars from APOGEE.

\begin{figure*}[t!]
\centering
\includegraphics[width=0.85\textwidth]{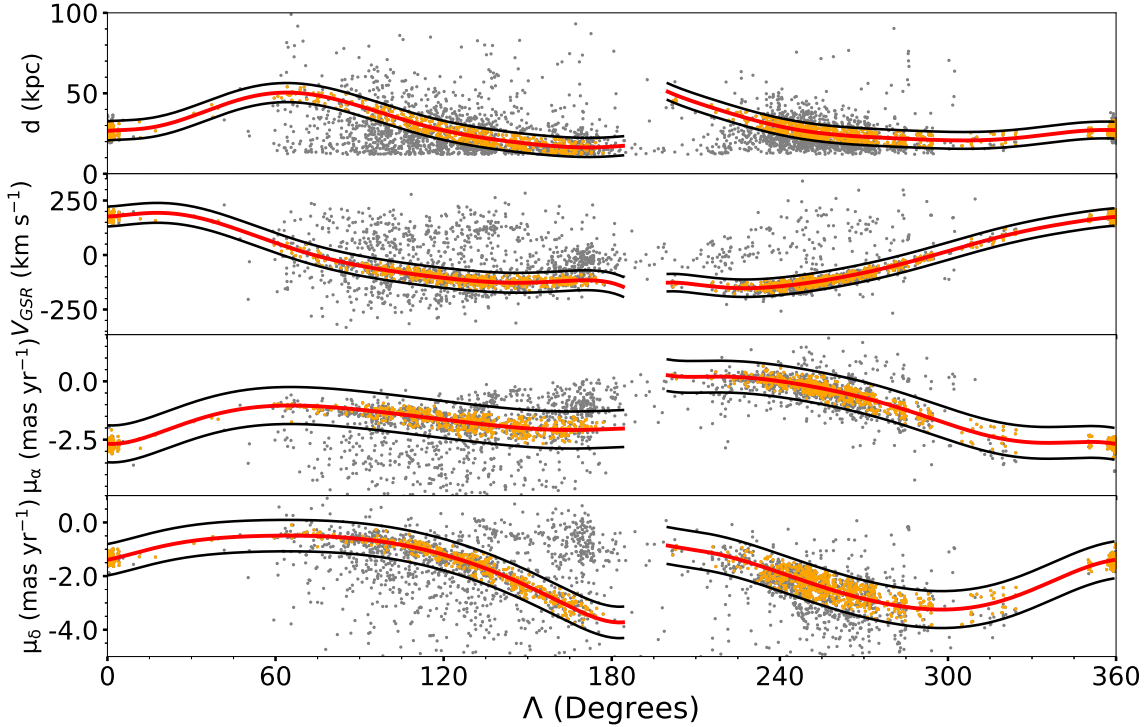}
\caption{Distribution of heliocentric distance ($d$), $V_{\rm GSR}$, $\mu_{\alpha}$,
and $\mu_{\delta}$ of the Sgr candidates, as a function of the Sgr stream
longitude ($\Lambda$). The gray dots represent the stars in $-20^{\circ} < B < 15^{\circ}$,
where $B$ is the Sgr stream latitude,
$d > 12$ kpc, and $L_Y < -0.3L_Z - 2.5~\times$ 10$^{3}$\ kpc \kms.
The red line in each panel is the fiducial used to select the likely Sgr member stars,
derived by a polynomial fit to the Sgr members in the catalog provided by V21.
The black line delineates the $\pm$3$\sigma$ curve from the fiducial, except for the distance
panel, considering the photometric distance error, which is set to
$\pm$4$\sigma$ away from the fiducial. The orange dots
indicate the stars within the black curves in all four panels. The cluster around
$\Lambda$ = 0$^{\circ}$ and 360$^{\circ}$ is the Sgr core (SC). The stars with $\Lambda$ $<$ 180$^{\circ}$
belong to the leading arm (LA), while the ones with $\Lambda$ $>$ 200$^{\circ}$ belong to the trailing arm (TA).}
\label{figure1}
\end{figure*}

\subsection{Space Velocity and Orbital Parameters} \label{sec2.2}

In order to calculate the orbital parameters of our program stars, we employed a St$\ddot{a}$ckel-type
potential used by numerous previous studies \citep{cb2000,kim2019,kim2021,kim2023,lee2019,lee2023}.
We adopted a local standard of rest velocity ($V_{\rm LSR}$) = 236 \kms\ \citep{kawata2019},
a solar position of $R_\odot$ = 8.2 kpc \citep{bhg2016}, $Z_\odot$ = 20.8 pc \citep{bb2019},
and a solar peculiar motion
of ($U$, $V$, $W$)$_\odot$ = (--11.10, 12.24, 7.25) \kms\ \citep{schonrich2010}.
We also computed the maximum (\rmax) and minimum (\rmin) distances from the Galactic center,
orbital eccentricity ($e$), defined by (\rmax\ -- \rmin)/(\rmax\ + \rmin), angular momentum ($L$),
and energy ($E$) for each star. The Galactic standard radial velocity ($V_{\rm GSR}$) was calculated by
$V_{\rm GSR}= V_{\rm H}-U_\odot{\rm cos}(l){\rm cos}(b)+(V_{\rm LSR}+V_\odot){\rm sin}(l){\rm cos}(b)+W_\odot{\rm sin}(b)$,
where $V_{\rm H}$ is the heliocentric radial velocity. {The uncertainty
of each computed quantity was estimated by incorporating the errors in the distance, proper motions,
and RV for each star.}

{We repeated the same calculations for
the simulated Sgr stars under the influence of the MW interacting with the LMC,
using the coordinates, distances, proper motions, and radial velocities
provided by V21\footnote[5]{Model data can be downloaded from \url{https://zenodo.org/record/4038141}.}.
We use our calculated eccentricity, angular momentum, and energy for these model stars
to compare with our Sgr member stars from the RGB/MS and RR Lyrae samples, as described in Section \ref{sec4}.}

\section{Selection of Sgr Stellar Stream Stars} \label{sec3}

\subsection{The RGB and MS Samples} \label{sec3.1}

The success of this study depends on how well we identify genuine members of the Sgr stellar stream.
Because identifying stellar debris in a coordinate
system aligned with a stellar stream can improve its characterization \citep{majewski2003,belokurov2014},
we introduced the Sgr stream coordinates ($\Lambda, B$) to select the members of the Sgr stream,
and converted the equatorial coordinates ($\alpha, \delta$) to Sgr stream coordinates using the
prescription of \citet{belokurov2014}. In these coordinates, the Sgr longitude ($\Lambda$) increases
in the direction of the Sgr core motion, and the Sgr latitude ($B$) points to the North Galactic Pole.
We made use of the positional and kinematic information with respect to the Sgr orbital plane,
and devised the following two-step selection criteria based on the Sgr members provided by V21
to reduce contamination.

\begin{enumerate}
    \item Based on the spatial distribution and distances of the Sgr stream stars, we initially select stars that are within --20$^{\circ} < B < 15^{\circ}$ and located at  heliocentric distance larger than $d =$ 12 kpc \citep{belokurov2014,ibata2020}. Then, we apply a broad kinematic cut of \ly\ $<$ --0.3\lz\ -- 2.5$\times$10$^{3}$\ kpc\ \kms\ \citep{li2019,jonsson2020,naidu2020}. The gray dots in Figure \ref{figure1} indicate the stars that pass these initial cuts. These stars exhibit a broad scatter in all four panels, indicating that they are not completely free of contamination from non-Sgr stars.

    \item In the second step, we employ the Sgr member catalog provided by V21 to further reduce the halo interlopers. The catalog includes 55,192 Sgr stars selected based on $\mid$$B$$\mid$ $<$ 10$^{\circ}$. To enhance the reliability of our Sgr selection, we acquire proper motion values of the catalog by matching with Gaia EDR3. We then consider the distribution of the Sgr candidates in four parameter spaces (distance, \vgsr, $\mu_{\alpha}$, and $\mu_{\delta}$), as a function of the stream longitude ($\Lambda$), as shown in Figure \ref{figure1}. We then carry out a polynomial fit in each parameter to obtain a fiducial line as a function of $\Lambda$. Each fiducial is indicated with a red line in Figure \ref{figure1}. The black lines in Figure \ref{figure1} are located at $\pm$3$\sigma$ from the fiducial line, except for the distance, which is at $\pm$4$\sigma$ from the fiducial line, considering the relatively large error of the photometric-based distance estimates of our program stars. To ensure that genuine Sgr member stars are chosen, we select objects that reside within the black curves in the four regions. These stars are our final Sgr stream and Sgr core (SC) samples, marked with orange colors in Figure \ref{figure1}. The stars with $\Lambda$ $<$ 180$^{\circ}$ belong to the leading arm (LA), while the ones with $\Lambda$ $>$ 200$^{\circ}$ to the trailing arm (TA).
    \end{enumerate}

{Following the above selection procedure, we identified
2380 RGB stars (1369 from APOGEE, 817 from SDSS (primarily SEGUE), and 194 from LAMOST)
and 37 MS stars, which come from only SDSS,
resulting in a total number of 2417 RGB/MS stars identified as likely Sgr members.
Among them, the LA comprises 555 RGB and 23 MS stars, with 662 RGB and 14 MS stars for the
TA. We have only RGB stars available from APOGEE for the SC. Consequently, the impact of the
MS stars on our analysis and interpretation is very minimal, even if there exists
any systematics in chemical abundances and orbital parameters between the RGB and MS stars.}

{Because the Sgr stars are mostly located in the distant halo,
we further checked the distance uncertainty for the selected Sgr members
after separating them into different luminosity classes and survey data.
By comparing the stars in common between SDSS and Gaia EDR3, we obtained
relative uncertainties of 22$\pm$1\% and 20$\pm$1\% for SDSS RGB and SDSS MS
stars, respectively. During this process, we adjusted for the reported zero-point offset of --0.017 mas in
Gaia EDR3 \citep{lindegren2021}, and considered only the stars
with relative parallax errors smaller than 10\%. We estimated the distance
uncertainties for APOGEE and LAMOST RGB stars by taking a median value
from the distribution of their quoted errors in their catalog. They are 27$\pm$5\%
and 22$\pm$8\% for APOGEE and LAMOST RGB stars, respectively.}

\begin{figure*}[!t]
\centering
\includegraphics[width=0.85\textwidth]{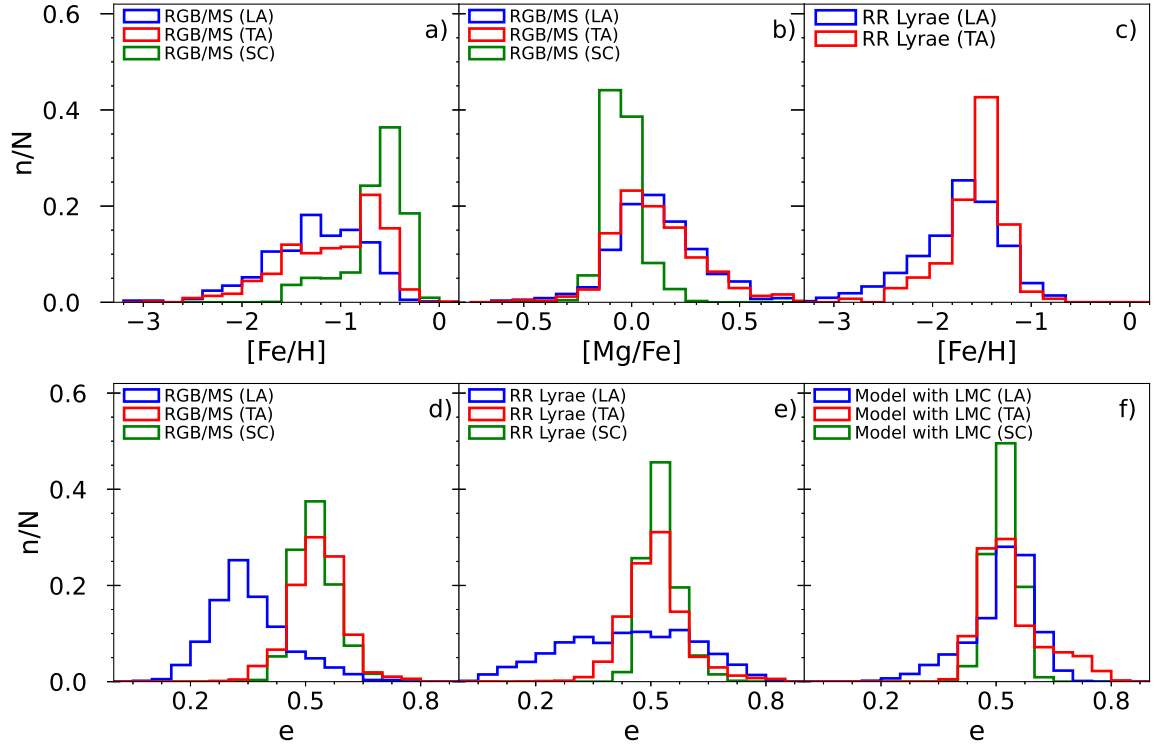}
\caption{{Top panels: Distribution of [Fe/H] (left) and [Mg/Fe] (middle)
for RGB/MS stars and [Fe/H] (right) for RR Lyrae stars. The metallicity information
for the RR Lyrae stars was obtained
by cross-matching the RR Lyrae catalog of \citet{ramos2020} with SDSS/LAMOST data.
The blue, red, and green histograms in each panel indicate
the LA, TA, and SC, respectively. Each
histogram is normalized by the total number of stars in each category.
Note that there are no [Mg/Fe] distributions for the RR Lyrae stars and no SC histogram in the right
panel, because there exist no SC stars in the SDSS/LAMOST data.
Bottom panels: Same as in the top panels, but for the distribution of eccentricity for
the RGB/MS (left), RR Lyrae (middle), and simulated Sgr stars with the LMC (right) from V21. We note
the large fraction of low-$e$ stars in the LA, and a similar behavior is observed in the model as well.}}
\label{figure2}
\end{figure*}

\subsection{The RR Lyrae Sample}\label{sec3.2}

One of the fundamental difficulties for study of the Sgr stream is its large distance
from the Sun ($\geq$ 10 kpc), and an uncertain kinematic analysis
due to the small number of stars close to the SC, along with distance errors that become
increasingly larger with increasing distance from the Sun.
This challenge can be partially rectified by using old stellar populations of similar
luminosity, such as blue horizontal-branch and RR Lyrae stars
\citep[e.g.,][]{belokurov2014,hernitschek2017,ramos2020}.
Hence, to increase the reliability of the dynamical analysis of our RGB/MS Sgr member
stars, we introduced a large number of RR Lyrae stars assembled by \citet{ramos2020}, which
have more accurate distance estimates, and carried out their dynamical analysis as well.
This sample includes about 11,700 RR Lyrae in the Sgr stream and core, selected based on the positions,
distances, and proper motions from
PanSTARRS1 \citep{chambers2016} and Gaia DR2 \citep{hernitschek2017}.
We gathered $\sim$ 7900 RR Lyrae stars that meet the selection criteria outlined 
in Section \ref{sec3.1}. See below how we obtained their RVs and orbital parameters.

One shortcoming of this sample is the lack of the RV information, which is indispensable to
derive the orbital parameters. In order to overcome this deficiency and retrieve the RV
information, we have devised the following procedure using a Monte Carlo (MC) simulation.

If the selected RR Lyrae stars are indeed members of the Sgr stream and/or core, their \vgsr\
values should fall inside the range indicated by the black lines in Figure \ref{figure1}.
On that basis, we attempted to recover the RV information using the \vgsr\ fitting line
and $\sigma$ values derived in Figure \ref{figure1}. To check the reliability of this
approach, we first tested this procedure with the 2417 Sgr member stars of the RGB/MS
sample that we obtained in Section \ref{sec3.1}. To begin, we assumed that there was
no RV for this sample of stars; hence, no \vgsr\ value. Then we employed the fiducial
line (red line) of \vgsr\ and $\sigma$ as estimates of the mean and standard deviation
of the \vgsr\ at a given $\Lambda$, obtained in the second panel of Figure \ref{figure1}.
Then, by assuming a Gaussian error distribution for \vgsr, we randomly drew a \vgsr\
value and assigned it to each star at
a specific $\Lambda$. We repeated this process 1000 times
to create 1000 simulated values of \vgsr\ for each object.
Finally, We calculated a mean value of \vgsr\ and plugged it to the \vgsr\ equation used in
Section \ref{sec2} to retrieve the RV value required to calculate the orbital parameters,
including the eccentricity.

A comparison of the retrieved RV obtained by the above method with
the observed RV for our RGB/MS stars yielded a mean offset of 0.4 \kms\
with a standard deviation of 19.0 \kms, indicating that the retrieved RV
from the MC simulation remains within 2$\sigma$ of the
quoted error value ($\sim$ 10 \kms). In addition, a comparison
of the derived eccentricity revealed a mean difference of only $\Delta\,(e) = 0.02$
with a standard deviation of $\sigma\,(e) = 0.04$.
This exercise confirms the robustness of our approach to recover the RV information for the
RR Lyrae stars. We applied the same method to the selected Sgr members of RR Lyrae stars
to assign the RV, and calculated their eccentricity, angular momentum, and energy.

\begin{figure*}
\centering
\includegraphics[width=0.95\textwidth]{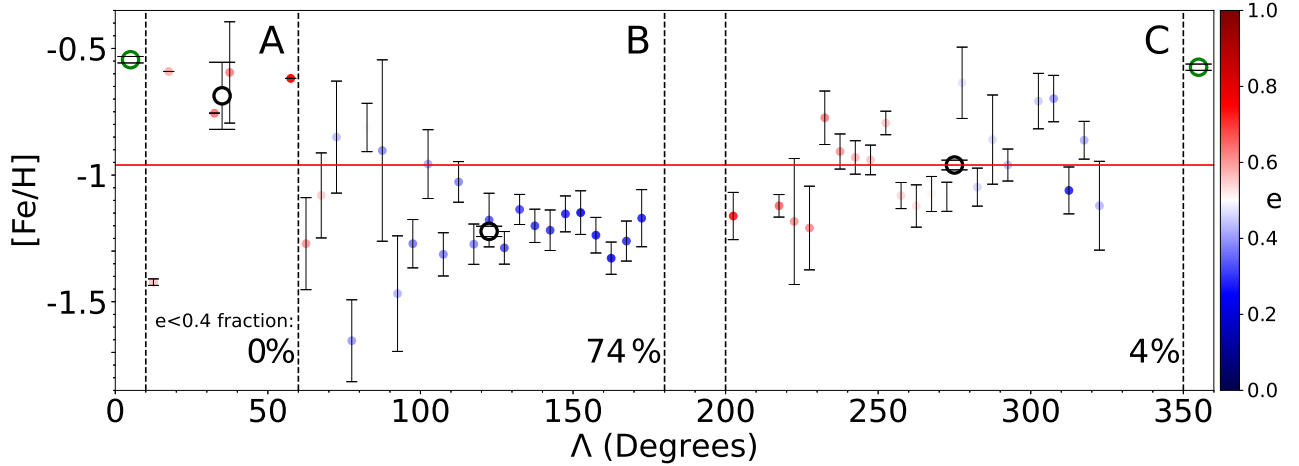}
\caption{Metallicity profile of three regions (A, B, and C), as a function of $\Lambda$
for the RGB/MS stars. Each point is a median value in a bin size of $\Lambda$ = 5$^{\circ}$;
the error bar is the standard deviation. {The color code represents the
median eccentricity in each bin, and its scale is shown in the right color bar.}
The large open circle represents the median metallicity
of the A, B, and C regions. Note that the LA is divided into two regions:
metal rich (A) and metal poor (B). The red horizontal lines indicates the median [Fe/H] of the TA.
At the bottom of the figure, the fraction of stars with $e < 0.4$ in each region is listed.}
\label{figure3}
\end{figure*}

\section{Results and Discussion} \label{sec4}

\subsection{Chemical Properties} \label{sec4.1}

Panel (a) of Figure \ref{figure2} exhibits the metallicity distribution
function (MDF) of our Sgr stream and Sgr core members of our RGB/MS sample, while panel (b)
is for the [Mg/Fe] distribution. The LA, TA, and SC members are represented by blue, red,
and green respectively. The MDFs indicate that the LA and TA have a broad
distribution, with the former being overall more metal poor than the latter, while
the SC is dominated by more metal-rich stars with a narrow peak. {The mean metallicity of
the LA is $\langle$[Fe/H]$_{\rm LA}$$\rangle$ = --1.14 $\pm$ 0.12 (sys) $\pm$ 0.03 (ran),
where the systematic error (sys) was derived by the standard deviation of the median metallicities
of APOGEE RGB, SDSS RGB, LAMOST RGB, and SDSS MS stars in the LA.
The random uncertainty (ran) was computed by error propagation of the available random errors
of the four subsamples. Similarly, the mean metallicities of the TA and SC
are $\langle$[Fe/H]$_{\rm TA}$$\rangle$ = --0.92 $\pm$ 0.07 (sys) $\pm$ 0.03 (ran)
and $\langle$[Fe/H]$_{\rm SC}$$\rangle$ = --0.56 $\pm$ 0.01 (ran), respectively.
Note that as we have only the giant stars from APOGEE in the SC,
we did not estimate the systematic error.} Our derived mean values
agree well with recent studies \citep[e.g.,][]{hayes2020,ramos2022,limberg2023}.

Panel (c) of Figure \ref{figure2} shows the MDF of the TA and LA
for the RR Lyrae stars. Although the peak of the MDFs for the LA and TA occurs at relatively lower
metallicity (around [Fe/H] = --1.68 and --1.50, respectively) than the RGB/MS sample,
due to the nature of their old stellar population, we clearly notice the systematic
difference in the MDFs between the LA and TA.

Panel (b) of Figure \ref{figure2} reveals that, while the SC has a narrow symmetric distribution
of [Mg/Fe], the distribution of the LA and TA is skewed to higher [Mg/Fe].
Quantitatively, {we obtained mean [Mg/Fe] values of +0.08 $\pm$ 0.07 (sys) $\pm$ 0.07 (ran),
+0.08 $\pm$ 0.07 (sys) $\pm$ 0.06 (ran), and --0.05 $\pm$ 0.02 (ran) for the LA, TA, and SC,
respectively. The systematic and random errors were calculated in the same way as for [Fe/H].}
\citet{hayes2020} found that the median [Mg/Fe] ratios of the LA, TA, and SC are +0.03, --0.01, and --0.03,
respectively, while \citet{ramos2022} obtained a mean value of \afe\ = +0.07 and +0.04 for the LA
and TA, respectively. These estimates are in good agreement with ours within the errors.

The chemical contrasts between the LA, TA, and SC suggest that the Sgr progenitor
had a different star-formation history in its center and outskirts, with a higher rate of
star formation in its center and a lower rate in its outskirts, establishing a radial abundance gradient.
According to the N-body simulation of the Sgr tidal disruption
by \citet{law2010}, the TA stars were stripped within the past $\sim$ 0.7 -- 3.0 Gyr,
while the LA stars were stripped $\sim$ 2.7 -- 5.0 Gyr ago.
Due to the outside-in nature of tidal stripping acting on a Sgr progenitor
with abundance gradients \citep[e.g.,][]{law2010,carlin2018,hayes2020}, the stars
in the LA, which were stripped first, are more metal poor than the ones in the TA.
The higher [Mg/Fe] implies that the stars in the Sgr stream formed
more rapidly, likely over a shorter timescale and with a higher rate of
star formation \citep{hayes2020}. In line with this,
\citet{hayes2020} found that their dynamically younger TA sample
falls between the LA and SC in both the [Fe/H] and $\alpha$-element abundances,
suggesting that both gradients were first established in the Sgr progenitor.

\subsection{Dynamical Properties} \label{sec4.2}

Panel (d) of Figure \ref{figure2} shows the eccentricity ($e$) distribution of the LA (blue),
TA (red), and SC (green) for our RGB/MS sample. One striking feature is that, although there
is an overlapping region, the $e$-distribution of the LA is completely different from that of
the TA and SC. The TA and SC exhibit very similar distributions to each other, biased toward
higher eccentricity ($e > 0.4$). {We found median
eccentricities of 0.34 $\pm$ 0.06, 0.55 $\pm$ 0.05, and 0.52 $\pm$ 0.08 for LA, TA, and SC,
respectively. Separating into the RGB and MS subsamples yielded 0.34 $\pm$ 0.06
for the RGB and 0.33 $\pm$ 0.08 for the MS in the LA, and
0.53$\pm$0.04 for the RGB and 0.56 $\pm$ 0.09 for the MS in the TA.
The uncertainties were derived from values of eccentricities created
1000 times by a MC method, after incorporating the uncertainties
in the distance, proper motions, and RV when calculating the eccentricity.
In addition, we checked the scatter of the eccentricity uncertainties to ensure that
the $e$-distribution for the RGB/MS sample is not an artifact generated
by errors in the RV, distance, and proper motions.
We obtained a dispersion of $0.06^{+0.06}_{-0.02}$ for the RGB
and $0.08^{+0.03}_{-0.01}$ for the MS in the LA, while $0.04^{+0.05}_{-0.01}$ for the RGB
and $0.09^{+0.01}_{-0.01}$ for the MS in the TA. For the SC, which includes only RGB stars,
the scatter is $0.08^{+0.01}_{-0.01}$. The errors were taken at 34\% to
the left and right from the median in each uncertainty distribution.
We note the consistent behavior without large dispersions in the two different samples,
although the MS sample is too small in each arm to compare to the RGB sample as mentioned in
Section \ref{sec3.1}. Nonetheless, this proves that the low-$e$ population in the LA indeed exists.
To our knowledge, this is the first time that a larger fraction of low-$e$ stars is found in
the LA compared with the TA and SC.}

Similarly, panel (e) of Figure \ref{figure2} exhibits the eccentricity
distribution of the Sgr RR Lyrae stars. As noted in the RGB/MS sample,
the TA and SC show a narrow distribution, with a peak of $e$ $\sim$ 0.5.
In comparison, the LA exhibits a much broader $e$-distribution than that of our RGB/MS sample,
ranging from 0.1 to 0.8 due to the large distance coverage than the RGB/MS sample.
Nonetheless, we clearly observe the low-$e$ population ($e < 0.4$),
confirming the existence of the low-$e$ RGB/MS stars in the LA. {The
small eccentricity error and its small dispersion of $0.05^{+0.07}_{-0.01}$ for
our RR Lyrae stars also support the presence of the low-$e$ population.}

Given that the stellar stream stripped off from a dwarf galaxy retains its progenitor's dynamical
properties such as eccentricity \citep{refiorentin2015,amorisco2017,mackereth2019},
one can expect that the LA, TA, and SC should have similar $e$-distributions.
Thus, our result highlights an interesting new feature.

To further investigate the disparate $e$-distribution of the LA, we plot
the metallicity profile of the Sgr members as a function of the stream
longitude ($\Lambda$) for the RGB/MS sample in Figure \ref{figure3}.
We excluded the RR Lyrae sample, because there is no metallicity information for the Sgr core stars.
Note that the LA region is divided into A (metal-rich)
and B (metal-poor) regions. The color code indicates the median
eccentricity. The fraction of stars with $e <$ 0.4 is listed in each region as well.
Despite a relatively large scatter in each region,
the figure clearly suggests that regions A and C have relatively
higher metallicity than that of B; closer to the SC (large green
circles), the metallicity becomes higher, implying some level of metallicity gradient
along each region, as other recent studies have reported \citep[e.g.,][]{gibbons2017,hayes2020,ramos2022},
We also observe that the low-$e$ fraction for the regions A and C is almost negligible,
whereas the B region has a large portion of stars with $e < 0.4$ (74\%).
{Additionally, we clearly see from the color code that the low-$e$ stars in the LA
are dominated by relatively metal-poor stars, located farther away from
the SC. Taking the metallicity as a proxy for time,
the older (metal-poor) population, which was stripped off first, possesses lower eccentricity. On
the other hand, the TA stars exhibit an opposite trend; the (metal-poor) stars farther away from the SC
exhibit higher eccentricity.} These distinct features between the LA and TA stars unambiguously indicate
a mixture of stars that have experienced different chemical and dynamical evolutionary histories.

\begin{figure*}
\centering
\includegraphics[width=0.8\textwidth]{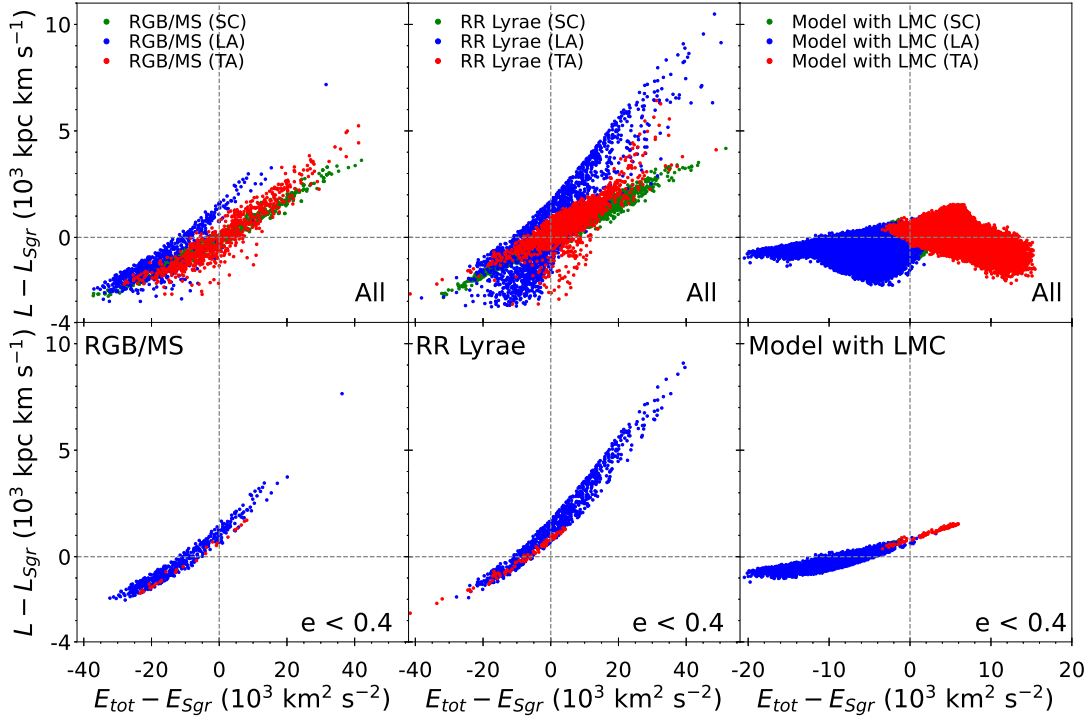}
\caption{{Angular momentum--energy ($L-E$) diagram for the RGB/MS stars (left panels),
the RR Lyrae stars (middle panels), and a Sgr model with LMC (right panels). The bottom
panels are the stars with $e < 0.4$. Both $L$ and $E$ quantities are re-scaled to have zero
at the location of the Sgr core as in V21. The color code is the same as in
Figure \ref{figure2}. Note that as the SC stars in the right panel are buried in the LA and TA, they are not
clearly visible in the panel. We clearly see the offset of the LA from the TA and SC (top left),
and we do not observe many stars with $e < 0.4$ in the TA and SC  (bottom left).
The RR Lyrae sample (bottom middle) exhibits the same pattern, and qualitatively, a similar behavior
is observed in the Sgr model with the LMC perturbation adopted from V21 (bottom right).}}
\label{figure4}
\end{figure*}

\begin{figure}
\centering
\includegraphics[width=0.95\columnwidth]{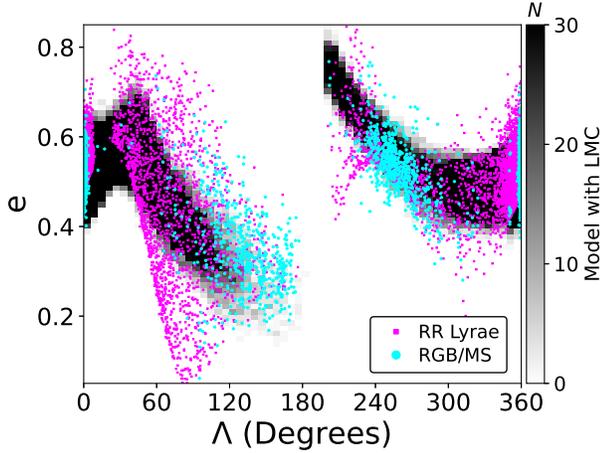}
\caption{Eccentricity distribution as a function of $\Lambda$. {The
cyan dots are the RGB/MS stars,
while the magenta points indicate the RR Lyrae stars. The black background
represents the number density produced by a Sgr model with LMC, which is adopted from V21.}
The cluster around $\Lambda$ $\sim$ 0$^{\circ}$ and 360$^{\circ}$
are the SC. We clearly observe that the eccentricity increases for the stars
closer to the SC for the LA, while the opposite behavior is seen for the TA; the farthest stars
from the SC have higher eccentricity.}
\label{figure5}
\end{figure}

\subsection{Signature of Dynamical Evolution of Low-$e$ Stars in the LA} \label{sec4.4}

We have found the $e$-distribution of the LA to be significantly different from that of the
TA and SC (Figure \ref{figure2}), having a large fraction of low-$e$ stars ($e < 0.4$)
in our RGB/MS sample. These stars are both more metal poor and located farther away from the SC,
as seen in Figure \ref{figure3}. This dynamical
dichotomy among the Sgr stream stars is also
reported in at least one other study. \citet{hasselquist2019} identified 35 Sgr member stars by
applying a clustering algorithm to the APOGEE chemical abundances,
and found that 26 of these stars have eccentricity in the range of 0.40 -- 0.85, and 9 likely Sgr
members that belong to the LA exhibit relatively lower $e$ ($<$ 0.4) than the Sgr main body.
These characteristics qualitatively agree with our findings. However, they did not give
much emphasis to these low-$e$ stars. They also noted that the TA stars exhibit a range
of \rmax\ = 60 -- 120 kpc, while a range of \rmax\ = 40 -- 60 kpc was found for the LA debris.
Because one expects that a stellar stream stripped
from a disrupted dwarf galaxy would follow similar dynamical
characteristics \citep[e.g.,][]{johnston1996,vandenbosch1999,refiorentin2015,amorisco2017,mackereth2019},
it is challenging to explain the contrasting $e$-behavior of the LA with
respect to that of the TA and SC.

Then, where did the low-eccentricity stars in the LA come from ? One might simply think
that they are interlopers, perhaps from another dwarf galaxy that
happened to be in the right place and have the right chemistry, similar to
the LA of the Sgr stream. However, as a different possible progenitor associated with
the low-$e$ stars of the LA has not been identified, this assertion appears unlikely.
It is more plausible to conjecture that the LA stars were stripped
off in the early history of the Sgr disruption, as many simulations
have suggested \citep{law2016,ibata2020,ramos2020,vasiliev2021},
and that they have been dynamically perturbed in such a way to alter
their orbits and become more circular than the TA and the SC.

{Using a catalog of the Sgr RGB stream members, V21 reported
a clear misalignment between the Sgr stream track and
the motion of the LA stars, after correcting for the Solar reflex motion
in their observed proper motions, suggesting that the time-varying potential of the MW
may perturb the Sgr's stream to alter its orbit. In
fact, V21 carried out Sgr simulations that
consider the perturbation of the MW caused by the LMC, and found that, without the LMC,
the Sgr stream stars would have larger apo-Galactic distances and higher eccentricity,
indicating that the presence of the LMC causes the decrease of the initial eccentricity
of the Sgr orbit. The LA is the most perturbed in their
simulations. Consequently, the low-$e$ stars we detected in the LA
can be qualitatively explained by their prediction.
It is important to recognize that the MW's reflex motion has a greater effect
on the Sgr stream than for the LMC itself, given that the LA is located in
the Northern Galactic Hemisphere, while the LMC is passing by in the Southern
Hemisphere.}

{The above claim can be confirmed in panel (f) of Figure \ref{figure2},
which exhibits the $e$-distribution of the simulated Sgr stars perturbed by the MW interacting
with the LMC. We grouped the model data into the LA, TA, and SC, same as in the RGB/MS
sample. Even though the fraction is relatively smaller
than that of the RGB/MS sample, we can identify the low-$e$ population ($<$ 0.4).
The reason that our RGB/MS stars have more low-$e$ stars is the lack
of the RGB/MS stars in the LA region, which are close to the SC (see Figure \ref{figure5}).
In fact, our RR Lyrae sample, which covers much larger distances and includes
more stars close to the SC, exhibits a lower fraction of low-$e$ stars in the LA (see panel (e) of
Figure \ref{figure2}) than the RGB/MS sample, confirming our claim above.}

{The different behavior in the $e$-distribution
should be reflected in the \rmax\ distribution as well. Our RGB/MS sample has a mean \rmax\ of 33.6 $\pm$ 0.6 kpc
for the LA and 53.4 $\pm$ 1.0 kpc for the TA, thus the LA has the shorter \rmax,
as in other studies. However, we obtained a mean \rmax\ of 33.9 $\pm$ 0.8 kpc
for the low-$e$ stars in the LA, unlike the expected shorter \rmax.
As mentioned above, this is because the LA stars in our RGB/MS sample are
dominated by the low-$e$ stars that are farther away from the SC. However, the RR Lyrae
stars, which contain more high-$e$ stars close to the SC, yielded a mean
\rmax of 56.6 $\pm$ 0.3 kpc, 49.2 $\pm$ 0.4 kpc, and 47.1 $\pm$ 0.7 kpc for the TA, LA,
and LA stars with $e < 0.4$, respectively, resulting in a smaller \rmax\ for the low-$e$ stars
in the LA, as expected. The Sgr model with perturbation from the LMC has a more uniform
stellar distribution, and clearly suggests a shorter distance of \rmax;
we derived a mean \rmax\ distance of 45.3 kpc and 37.6 for the LA and
the LA stars with $e < 0.4$, respectively. Therefore, we can
interpret that the \rmax\ distance of low-$e$ stars in the LA
became shorter due to the influence of the MW's reflex motion.}

{To investigate the connection of the low-$e$ stars with the perturbation
of the LMC more closely, we used the RGB/MS, RR Lyrae, and Sgr model samples
to produced the angular momentum--energy ($L-E$) diagram in Figure \ref{figure4},
similar to figure 11 of V21. The $L$ and $E$ quantities
are re-scaled to be zero at the Sgr core in this Figure. As V21 well
illustrated, models of tidal-tail formation \citep[e.g.,][]{helmi2000,eyre2011,gibbons2014}
produce trailing and leading arms that have distinct patterns in the $L-E$ plane, in that they
are shifted in opposite directions from their progenitor dwarf. That is, the trailing
stars have a higher energy and angular momentum, while the leading stars possess lower energy and angular
momentum compared to their progenitor (see panel (d) of figure 11 of V21).
However, we do not observe this behavior in the top-left panel of
Figure \ref{figure4}; rather, the LA occupies a slightly higher angular momentum region than
the SC at a given $E$, while the TA and SC are overlapped with one another.
This behavior is, to some degree, in line with the right panel of this Figure, with no large offset
seen in the angular momentum among LA, TA, and SC, which was reproduced by adopting the model
data of V21 that consider the time-dependent potential of the MW interacting with the LMC.
Note that we recalculated $L$ and $E$ for each star, using
the St$\ddot{a}$ckel-type potential, as mentioned in Section \ref{sec2.2}.
In the V21 simulations, the LMC perturbation is expected to
cause the angular momentum (eccentricity) of the LA to increase (decrease) and the
angular momentum (eccentricity) of the TA to decrease (increase).}

{The impact of the LMC perturbation is more clearly seen
in the Sgr model data, which only show the stars with $e < 0.4$ (bottom-right panel of Figre \ref{figure4}).
It shows some fraction of stars with angular momentum differences larger than
--1000 kpc \kms\ in the LA, whereas the TA and SC stars are almost removed.
Similarly, there is a lack of the SC and TA stars in the RGB/MS sample (bottom-left panel).
Consequently, the top-left panel of Figure \ref{figure4} signals a
significant evolution of the angular momentum of the LA after its stripping.}

{Our RR Lyrae sample also exhibits similar features, as seen in the middle panels
of Figure \ref{figure4}, but thanks to the greater distance coverage
and a larger sample size, we observe a larger fraction of stars with higher $L$ and $E$ stars
than for the RGB/MS stars. As can be inferred from the bottom-middle panel, the stars with higher $L$ than
that of the SC sequence have low eccentricity. These low-$e$ stars
are located farther away from (closer to) the SC in the case of the LA (TA),
as can be appreciated from inspection of Figure \ref{figure5}, which exhibits
the eccentricity distribution as a function of $\Lambda$. Consequently,
our findings provide additional evidence for the influence
of the MW by the LMC on the dynamical evolution of the tidal debris of the Sgr stream,
as envisaged by V21.}

{The fact that the TA is overlapped with the SC in the $L-E$ plane suggests that
the TA is also affected by the perturbation. According to the simulation by V21,
the TA stars, which are located farther away from the SC,
experience more angular-momentum evolution due to disturbance by the LMC.
As a result, they possess lower angular momentum, hence higher eccentricity.
As evidence, we notice some fraction of high-$e$ stars in the TA in panel (f)
of Figure \ref{figure2}. Observationally, we can assess this behavior among the RR Lyrae stars (magenta)
in Figure \ref{figure5} as well. In this Figure, the TA stars farther away from
the SC ($200^{\circ} < \Lambda < 240^{\circ}$) have
more eccentric orbits (e.g., lower angular momentum), in contrast to the LA stars ($\Lambda < 180^{\circ}$).
The Sgr model (black background) with perturbation from the LMC confirms this pattern.
These observational features clearly present that the stars
close to the SC are less perturbed by the time-dependent potential
of the MW by the LMC.}

{To conclude, our findings and interpretation confirm the dynamical evolution
of the Sgr stream due to the reflex motion of the MW's center caused by
the gravitational influence of the LMC. However, from a chemical perspective, one caveat of this dynamical
interpretation, following the simulations of V21, is that if the LA and TA
were simultaneously stripped off, it is difficult to reconcile
the metallicity discrepancy between the LA and TA.}

\section{Conclusions} \label{sec5}

We have presented chemical and dynamical properties of the Sgr member stars.
The MDF of the LA is slightly shifted to a more metal-poor level
than that of the TA. {We obtained a mean value of
[Fe/H] = --1.14 $\pm$ 0.12 (sys) $\pm$ 0.03 (ran) for the LA,
and [Fe/H] = --0.92 $\pm$ 0.07 (sys) $\pm$ 0.03 (ran) for the TA.
The MDF of the SC is biased toward the more metal-rich region, yielding
a mean of [Fe/H] = --0.56 $\pm$ 0.01 (ran). We derived mean values of
[Mg/Fe] = +0.08 $\pm$ 0.07 (sys) $\pm$ 0.07 (ran), +0.08 $\pm$ 0.07 (sys) $\pm$ 0.06 (ran),
and --0.05 $\pm$ 0.02 (ran) for the LA, TA, and SC, respectively.
These chemical features of the Sgr members agree well with recent
studies \citep[e.g.,][]{hayes2020,ramos2022,limberg2023}.}

The chemical contrast among the LA, TA, and SC can be understood
by a radial metallicity gradient due to different
star-formation histories in the progenitor of the Sgr dwarf.
Because the old, metal-poor stars at the outer edge
of the Sgr progenitor were preferentially stripped off,
the stars in the LA, which were stripped first, are more metal poor
than the stars in the TA.

{One striking finding in this study is the relatively
larger fraction of low-$e$ ($<$ 0.4) stars in the LA than in
the TA and SC. As one can expect that a stellar stream stripped off from a dwarf galaxy closely follows
the eccentricity of its progenitor, this behavior presents a challenge to understand.
We attempted to explain the disparate $e$-distribution of the LA
by appealing to time-dependent perturbations of the MW by the LMC, following
the demonstration of V21. According to their simulations,
the LA is perturbed the most, and the disturbance
from the LMC to the Sgr tidal stream causes the decrease of
eccentricity of the Sgr orbit, qualitatively in accordance with
the eccentricity behavior seen in our LA stars.}

The simulation by V21 also predicts that the TA stars,
which are located farther away from the Sgr core, experience more dynamical
evolution due to disturbance from the LMC. We confirm this by finding high-$e$ RR Lyrae
stars in the TA region that are more distant from the SC.

\begin{acknowledgments}
We thank an anonymous referee for a careful review of this paper,
which has improved the clarity of its presentation.
This work was supported by Chungnam National University.
T.C.B. acknowledges partial support for
this work from grant PHY 14-30152; Physics Frontier Center/JINA Center for the Evolution
of the Elements (JINA-CEE), awarded by the U.S. National Science Foundation,
and from OISE-1927130: The International Research Network for Nuclear Astrophysics (IReNA),
awarded by the U.S. National Science Foundation.


Funding for the Sloan Digital Sky Survey IV has been provided by the
Alfred P. Sloan Foundation, the U.S. Department of Energy Office of
Science, and the Participating Institutions.

SDSS-IV acknowledges support and resources from
the Center for High Performance Computing  at the University of Utah. The SDSS
website is www.sdss.org.

SDSS-IV is managed by the Astrophysical Research Consortium
for the Participating Institutions of the SDSS Collaboration including
the Brazilian Participation Group, the Carnegie Institution for Science,
Carnegie Mellon University, Center for Astrophysics | Harvard \&
Smithsonian, the Chilean Participation Group, the French Participation Group,
Instituto de Astrof\'isica de Canarias, The Johns Hopkins
University, Kavli Institute for the Physics and Mathematics of the
Universe (IPMU) / University of Tokyo, the Korean Participation Group,
Lawrence Berkeley National Laboratory, Leibniz Institut f\"ur Astrophysik
Potsdam (AIP), Max-Planck-Institut f\"ur Astronomie (MPIA Heidelberg),
Max-Planck-Institut f\"ur Astrophysik (MPA Garching),
Max-Planck-Institut f\"ur Extraterrestrische Physik (MPE),
National Astronomical Observatories of China, New Mexico State University,
New York University, University of Notre Dame, Observat\'ario
Nacional / MCTI, The Ohio State University, Pennsylvania State
University, Shanghai Astronomical Observatory, United
Kingdom Participation Group, Universidad Nacional Aut\'onoma
de M\'exico, University of Arizona, University of Colorado Boulder,
University of Oxford, University of Portsmouth, University of Utah,
University of Virginia, University of Washington, University of
Wisconsin, Vanderbilt University, and Yale University.

The Guoshoujing Telescope (the Large Sky Area Multi-
Object Fiber Spectroscopic Telescope, LAMOST) is a National
Major Scientific Project which is built by the Chinese Academy
of Sciences, funded by the National Development and Reform
Commission, and operated and managed by the National
Astronomical Observatories, Chinese Academy of Sciences.
\end{acknowledgments}

\bibliographystyle{aasjournal}

 \label{reference}

\end{document}